\documentstyle[12pt]{article}

\newcommand{\beq}{\begin{equation}}
\newcommand{\eeq}{\end{equation}}
\newcommand{\ben}{\begin{eqnarray}}
\newcommand{\een}{\end{eqnarray}}
\begin{document}
\begin{center}
{\large \bf A noncommutative anomaly through Seiberg-Witten map and \\ non-locally regularized BV quantization} \\

\vspace{1cm}

{\large Everton M. C. Abreu$^{a,b,c,}$\footnote{\noindent e-mail: evertonabreu@ufrrj.br} and Vahid Nikoofard$^{b,}$\footnote{\noindent e-mail: vahid@fisica.ufjf.br}}\\

%\end{center}
\vspace{1cm}

${}^{a}$Grupo de F\' isica Te\'orica e Matem\'atica F\' isica, Departamento de F\'{\i}sica,
Universidade Federal Rural do Rio de Janeiro\\
BR 465-07, 23890-971, Serop\'edica, Rio de Janeiro, Brazil\\
${}^{b}$Departamento de F\'{\i}sica, ICE, Universidade Federal de Juiz de Fora,\\
36036-330, Juiz de Fora, MG, Brazil\\
${}^{c}$LAFEX, Centro Brasileiro de Pesquisas F\' isicas (CBPF), Rua Xavier Sigaud 150,\\
Urca, 22290-180, RJ, Brazil\\

\bigskip

\today

\end{center} 

\vspace{0.5cm}
\abstract

\noindent Anomalies are one essential concept for the renormalization of noncommutative (NC) gauge theories.  A NC space can be visualized as a deformation of the usual spacetime with the $\star$-product and can be constructed after the quantization of a given space with its symplectic structure.  The Seiberg-Witten (SW) map connects NC fields, transformations parameters and gauge potential to their commutative analogs.   In this work we used the SW map to calculate the NC version of the anomaly of the BV quantized
chiral Schwinger model with nonlocal regularization.

\vskip 3cm
\noindent PACS: 1.10.Nx; 11.10.Gh; 11.15.-q

\vfill\eject

\section{Introduction}

It is a well known fact that quantum field theory has its main basis in the 
principle of gauge symmetry \cite{iz}.  The gauge theory, constructed with the principle of gauge symmetry, encompassing the symmetries and their corresponding conservation laws, has underlying role in the description of the fundamental forces in nature.  Nevertheless, we have also to consider that a specific conservation law that is true in a certain classical theory, can be broken when the theory is quantized.   In this case we have what is known as an anomaly (for a review see \cite{bertlman}).  Anomalies have their importance in physics where they are needed to describe certain experimental facts, for example.  The anomaly cannot be considered just a perturbation effect, which results from the regularization of some divergent diagrams, it shows the deep laws of quantum physics.  So, as we can see, it is important to find methods to compute the anomaly by the quantized the primary theory.

After quantization, we have the classical dynamical variables of the theory becoming noncommuting operators.  This fact makes us to believe that the classical   manifold framework of spacetime at the quantum (Planck) scale should have some kind of noncommutative (NC) structure.  So, to deal with this quantum gravity theory we have to consider a quantum field theory beyond a structure that depends on locality.   To understand these and other issues, it is a common sense, nowadays, in theoretical physics that NC geometry have the proper, precise and rigorous formalism to accomplish this target (see \cite{NC} for reviews).

Concerning NC manifolds, the anomalies in gauge theories has been explored with some intensity.  It was realized the importance of the structure of anomalies in this scenario.
The map developed by Seiberg and Witten \cite{sw} establish the validity of classical gauge transformations for theoretical systems constructed in NC and ordinary commutative spacetimes.  Through this analysis, we have an alternative procedure to investigate NC gauge theories via their commutative analogs.

In this paper we compute the anomaly of the chiral Schwinger model (CSM) in a NC bi-dimensional spacetime manifold.  After reviewing the non-local regularization BV formalism, we have applied the SW map to obtain the NC version of the anomaly.  Although the CSM anomaly is a well known result in the current literature, its NC analog is a new one.

The method developed by Batalin-Vilkovisky (BV) \cite{BV} showed itself to
be a very powerful way to quantize the most difficult field theories.  
A two dimensional gauge theory, the string theory, is one of these examples.  
For a review see \cite{Jon,Gomis,Hen}.
The BV, or field-antifield, formalism provides, at Lagrangian level, a general 
framework for covariant path integral quantization of gauge theories.  
This formalism uses interesting mathematical objects like
a Poisson-like bracket (the antibracket), canonical 
transformations, ghosts for the BRST transformations, etc.  The most important 
object of this method at the classical level is an equation called 
classical master equation (CME).

It is important to say that the CSM has been
constructed and completely solved by Jackiw and Rajaraman \cite{JR}. 
The organization of the paper is as follows.  In section 2 we described the main relation between the SW maps and the anomaly. 
In section 3 a brief review of the field-antifield formalism and its regularization has been made.
The computation of the NC CSM anomaly at one-loop has been calculated in section 4.  The conclusions and final remarks were accomplished in section 5.

\section{Noncommutativity and the Seiberg-Witten map}

In few words we can say that the noncommutativity used follows the idea of the deformation of the Minkowsky space with a real antisymmetric and constant parameter $\theta^{\mu\nu}$ such that $[x^{\mu}, x^{\nu}]\,=\,x^{\mu} \star x^{\nu}\,-\,x^{\mu} \star x^{\nu}\,=\,i\theta^{\mu\nu}$, where the $\star$-product will be defined in a jiffy.  It is known also as the Moyal-Weyl product.

To obtain the SW maps we have to take two different limits of string theory.   Hence, we can define a map between Nc fields and theirs ordinary analogs.  A gauge equivalence relation can be written as

\beq
\label{aaaaa}
\hat{A}_{\mu}(A,\theta)\,+\,\hat{\delta}_{\hat{\lambda}}\hat{A}_{\mu}(A,\theta)\,=\,\hat{A}_{\mu}(A\,+\,\delta_{\alpha}\,A, \theta)
\eeq
where $\alpha$ and $A$ are the ordinary gauge parameter and gauge field respectively.  $\delta_{\alpha}$ is the ordinary gauge transformation,

\beq
\label{bbbbb}
\delta_{\alpha} A_{\mu}\,=\,\partial_{\mu} \alpha\,-\,i[A_{\mu}, \alpha]\,=\,D_{\mu} \alpha
\eeq

\noindent So,we can write (\ref{aaaaa}) as 

\ben
\label{ccccc}
\hat{\delta}_{\hat{\lambda}}\,A_{\mu}(A,\theta)\,&=&\,\hat{A}_{\mu}(A\,+\,\delta_{\alpha}A, \theta)\,-\,\hat{A}_{\mu} (A,\theta) \nonumber \\
&=&\delta_{\alpha}\,\hat{A}_{\mu} (A, \theta)
\een

\noindent It is important to say that the NC gauge field $\hat{A}$ and the NC gauge parameter $\hat{\lambda}$ can be defined obeying the following dependence,

\ben
\label{ddddd}
\hat{A}_{\mu}\,=\,\hat{A}_{\mu}\,(A,\theta) \nonumber \\
\hat{F}_{\mu\nu}\,=\,\hat{F}_{\mu\nu}\,(A,\theta) \\
\hat{\lambda}\,=\,\hat{\lambda}\,(\alpha,A,\theta) \nonumber
\een

Hence, we have to solve ({\ref{aaaaa}) simultaneosly for $\hat{A}_{\mu}$ and $\hat{\lambda}_{\lambda}$ which is a difficult task.  However, this difficulty can be solved by generalizing the ordinary gauge condition

\beq
\label{eeeee}
\delta_{\alpha}\,\delta_{\beta}\,-\,\delta_{\beta}\delta_{\alpha}\,=\,\delta_{-i[\alpha,\beta]}
\eeq
to the NC case

\beq
\label{fffff}
i\delta_{\alpha}\,\hat{\lambda}_{\beta}\,-\,i\delta_{\beta}\,\hat{\lambda}_{\alpha}\,-\,[\hat{\lambda}_{\alpha}, \hat{\lambda}_{\beta}]_{\star}\,=\,i\hat{\lambda}_{-i[\alpha,\beta]}
\eeq

\noindent where this equation focuses only on the parameter $\hat{\lambda}_{\alpha}$ \cite{jmssw} and the solutions can be computed order by order \cite{uy}.
To find the SW maps, one have to solve (\ref{fffff}) and (\ref{ccccc}), respectively, order by order in $\theta$.  

After that explanation we will explain the main steps of SW work \cite{sw} which objective was to construct a bridge between commutative and NC field.  Nevertheless, our aim is to use the SW map to calculate the NC version of CSM from its commutative analog, as we said above.

Let us begin with the well known maps in a $U(1)$ gauge theory given by

\beq
\label{a1}
\hat{A}_{mu}\,=\,A_{\mu}\,-\,\frac 12 \theta^{\alpha\beta}\,A_{\alpha}\,(\partial_{\beta}A_{\mu}\,+\,F_{\beta\mu})\,+\,O(\theta^2)
\eeq
\beq
\label{a2}
\hat{F}_{\mu\nu}\,=\,F_{\mu\nu}\,-\,
\theta^{\alpha\beta}\,(A_{\alpha}\partial_{\beta}F_{\mu\nu}\,+\,F_{\mu\alpha}F_{\beta\nu})\,+\,O(\theta^2)
\eeq
\beq
\label{a3}
\hat{\lambda}\,=\,\lambda\,-\,\frac 12 \theta^{\alpha\beta}A_{\alpha}\partial_{\beta}\lambda\,+\,O(\theta^2)
\eeq

\noindent where the hat indicates that the variable is NC.  It is easy to see that this map is a gauge equivalence between the NC gauge theory and its ordinary analog.  Also, the map (\ref{a2}) is a direct result from map (\ref{a1}) since we have also that

\beq
\label{a4}
\hat{F}_{\mu\nu}\,=\,\partial_{\mu}\hat{A}_{\nu}\,-\,\partial_{\nu}\hat{A}_{\mu}
\,-\,i[\hat{A}_{\mu},\hat{A}_{\nu}]_{\star}
\eeq

\noindent where $[\hat{A}_{\mu},\hat{A}_{\nu}]_{\star}\,=\,\hat{A}_{\mu} \star \hat{A}_{\nu}\,-\,\hat{A}_{\nu}\star \hat{A}_{\mu}$ and the $\star$-product is the so-called Moyal-Weyl product or star product which, for two fields $A(x)$ and $B(y)$ is given by

\beq
\label{a5}
\left. (A\star B)(x)\,=\,exp\left(\frac i2 \theta^{\alpha\beta}\partial^x_{\alpha}\,\partial^y_{\beta}\right)\,A(x)B(y)\right|_{x=y}
\eeq

\noindent So, $\hat{F}_{\mu\nu}$ in (\ref{a4}) can be written as,

\beq
\label{a6}
\hat{F}_{\mu\nu}\,=\,\partial_{\mu}\hat{A}_{\nu}\,-\,\partial_{\nu}\hat{A}_{\mu}
\,+\,\theta^{\alpha\beta}\partial_{\alpha}\hat{A}_{\mu} \partial_{\beta}\hat{A}_{\nu}\,+\,O(\theta^2)
\eeq

\noindent However, notice that, since we know that $F_{\mu\nu}=\partial_{\mu} A_{\nu}-\partial_{\nu} A_{\mu}$ is gauge invariant, we have that $\hat{F}_{\mu\nu}$ transforms covariantly under the star gauge transformation \cite{bk1,bk2} given by

\ben
\label{a7}
\hat{\delta}_{\hat{\lambda}}\,\hat{F}_{\mu\nu}\,&=&\,i [\hat{\lambda},\hat{F}_{\mu\nu}]_{\star} \nonumber \\
&=&\theta^{\alpha\beta}\,\partial_{\alpha} \hat{F}_{\mu\nu} \partial_{\beta} \hat{\lambda} \,+\,O(\theta^2)
\een

\noindent which stability is granted by the gauge transformations given in (\ref{a1})-(\ref{a3}) such that

\ben
\label{a8}
\hat{\delta}_{\hat{\lambda}}\,\hat{A}_{\mu}\,&=& \hat{D}_{\mu}\star \hat{\lambda} \,\equiv\,\partial_{\mu}\hat{\lambda}\,+\,
\,i [\hat{\lambda},\hat{A}_{\mu}]_{\star} \nonumber \\
&=&\partial_{\mu}\hat{\lambda}\,+\,\theta^{\alpha\beta}\,\partial_{\alpha} \hat{A}_{\mu} \partial_{\beta} \hat{\lambda} \,+\,O(\theta^2)
\een
and $\delta_{\lambda}A_{\mu}\,=\,\partial_{\mu} \lambda$.

To analyze NC gauge theories we have to make the transition between commutative and NC gauge theories.  For example, if we are treating a NC action defined by \cite{bk1,bk2}

\beq
\label{1A}
\hat{S}(\hat{A},\hat{\Psi})\,=\,-\frac 14 \int d^4 x \hat{F}_{\mu\nu}\star \hat{F}^{\mu\nu}\,+\,\hat{S}_M (\hat{\Psi},\hat{A})
\eeq

\noindent where the first term $(\hat{S}_{\hat{A}})$ is for the gauge field alone and the second one is for the charged matter fields $\hat{\Psi}$.  The equation of motion for $\hat{A}_{\mu}$ is

\beq
\label{11A}
\frac{\delta\hat{S}_{\hat{A}}}{\delta\hat{A}_{\mu}}\,=\,\hat{D}_{\nu}\star \hat{F}^{\nu\mu}\,=\,\hat{J}^{\mu}
\eeq

\noindent where

\beq
\label{12A}
\hat{J}^{\mu}\,=\,-\,\left.\frac{\delta\hat{S}_{M}}{\delta\hat{A}_{\mu}}\right|_{\hat{\Psi}}
\eeq

If we susbtitute the SW map (\ref{a1})-(\ref{a3}) into the action in (\ref{1A}) we will have the NC parameters within the commutative action as we can see in 

\beq
\label{A1}
\hat{S}(\hat{A},\hat{\Psi}) \rightarrow S(A,\Psi,\theta)\,=\,S_A (A,\theta)\,+\,S_M (A,\Psi,\theta)
\eeq
where
\beq
\label{A2}
S_A (A,\theta)\,=\, -\frac 14 \int d^4 x \left[F_{\mu\nu}^2\,+\,\theta^{\alpha\beta} F^{\mu\nu}(2F_{\mu\alpha}F_{\nu\beta}\,\frac 12 F_{\beta\alpha}F_{\mu\nu})\,+\,O(\theta^2) \right]
\eeq

\noindent From Eq. (\ref{A1}) we have that

\beq
\label{A3}
\frac{\delta S_{A}(A,\theta)}{\delta {A}_{\mu}}\,=J^{\mu}\,=\,-\frac{\delta S_{M}(A,\theta)}{\delta {A}_{\mu}}
\eeq

\noindent and, directly, we obtain the conservation law

\beq
\label{A4}
\partial_{\mu}\,J^{\mu}\,=\,0
\eeq

\noindent From (\ref{a2}) we can write,

\beq
\label{A5}
\hat{J}^{\mu}\,=\,J^{\mu}\,-\,\theta^{\alpha\beta} A_{\alpha} \partial_{\beta} J^{\mu}\,+\,O(\theta^2)
\eeq

Notice that in (\ref{A5}) we have the freedom of adding more $O(\theta)$ terms such that these extra terms are invariant under ordinary gauge transformation.  The most general expression is given by,

\beq
\label{A55}
\hat{J}^{\mu}\,=\,J^{\mu}\,-\,\theta^{\alpha\beta}A_{\alpha} \partial_{\beta} J^{\mu}\,+\,c_1\, \theta^{\mu\alpha} F_{\alpha\beta} J^{\beta}\,+\,c_2\, \theta^{\alpha\beta} F_{\alpha\beta} J^{\mu}\,+\,c_3\, \theta^{\alpha\beta} F_{\alpha}^{\mu} J_{\beta}\,+\,O(\theta^2)
\eeq

\noindent where $c_1 , c_2$ and $c_3$ are parameters to be determined \cite{bk1,bk2}.  For instance, for a simultaneous conservation we can write that

\beq
\label{A6}
\hat{D}_{\mu}\star \hat{J}^{\mu}\,=\,\partial_{\mu} J^{\mu}
\eeq

\noindent which fixes $c_1 = 2c_2 =1$ and $c_3 =0$, and 

\beq
\label{A7}
\hat{J}^{\mu}\,=\,J^{\mu}\,-\,\theta^{\alpha\beta} \partial_{\beta}(A_{\alpha}J^{\mu})\,+\,\theta^{\mu\alpha}F_{\alpha\beta}J^{\beta}\,+\,O(\theta^2)
\eeq

\noindent And using (\ref{a1}) and (\ref{A7}), the covariant divergence of $\hat{J}^{\mu}$,

\ben
\label{A19}
\hat{D}_{\mu}\star \hat{J^{\mu}}\,&=&\,\partial_{\mu}\,\hat{J}^{\mu}\,+\,i[\hat{J}^{\mu},\hat{A}_{\mu}]_{\star} \nonumber \\
&=&\partial_{\mu}\hat{J}^{\mu}\,-\,
\theta^{\alpha\beta}\partial_{\alpha}\hat{J}^{\mu}\partial_{\beta}\hat{A}_{\mu}
\,+\,O(\theta^2) \nonumber \\
&=&\partial_{\mu}{J}^{\mu}\,+\,
\theta^{\alpha\beta}\partial_{\alpha}(A_{\beta}\partial_{\mu}{J}^{\mu})
\,+\,O(\theta^2)
\een

\noindent and we see clearly in this expression that the covariant conservation of $\hat{J}^{\mu}$ relies on the conservation of $J^{\mu}$.

To attack the real issue here, the anomaly, we can use the study developed until now for the vector current to derive a map for the axial current.  As we know, at the quantum level, the axial currents are not conserved.  The well known ABJ current \cite{abj} is not modified by noncommutativity and is written as

\beq
\label{a5555}
\partial_{\mu}{J}_5^{\mu}\,=\,{\cal A}\,=\,\frac{1}{16\pi^2}\,\epsilon_{\mu\nu\lambda\rho}\,F^{\mu\nu}\,F^{\lambda\rho}
\eeq

\noindent and the anomaly in the NC manifold is given by what we saw above so that

\ben
\label{a6666}
\hat{\cal A}\,&=&\,\hat{D}_{\mu}\star \hat{J}_5^{\mu} \nonumber \\
&=&\frac{1}{16\pi^2}\,\epsilon_{\mu\nu\lambda\rho}\hat{F}^{\mu\nu}\star \hat{F}^{\lambda\rho}
\een

Finally, the cherished map for anomalies, obtained from (\ref{A19}), is given by,

\beq
\label{a7777}
\hat{{\cal A}}\,=\,{\cal A}\,+\,\theta^{\alpha\beta}\,\partial_{\alpha}(A_{\beta}{\cal A})\,+\,O(\theta^2)
\eeq

\noindent which was demonstrated to be valid \cite{bk2}, although derivative corrections are necessary at higher orders.  So, now we cab establish the axial current at the quantum level, so that

\beq
\label{a8888}
\hat{J}_5^{\mu}\,=\,{J}_5^{\mu}\,-\,\theta^{\alpha\beta}\,\partial(A_{\alpha}{J}_5^{\mu})
\,+\,\theta^{\mu\alpha}\,F_{\alpha\beta}\,J_5^{\beta}\,+\,O(\theta^2)\,\,,
\eeq
which can be used to investigate anomalous commutators in NC electrodynamics, for example \cite{bk1}.

\section{The Field-Antifield Formalism}

The basic idea of BV formalism is BRST invariance.  The ingredients are the fields 
$\Phi^{A}$, i.e., the classical fields of the theory, the 
ghosts, the auxiliary fields and their canonically conjugated antifields $\Phi^{*}_{A}$.  
With all this 
elements we construct the so called BV action.  At the classical level, 
the BV action becomes the classical action when all the antifields are put 
to be zero.  A gauge-fixed action can be obtained by a canonical 
transformation.  At this time we can say that the action is in a gauge-fixed 
basis.  The other way to fix the gauge is through the choice of a gauge 
fermion  and to make the antifields to be equal to the functional derivative of this 
fermion.  

The method can be applied to gauge theories which have an open 
algebra (the algebra of gauge transformations closes only on shell),
to closed algebras, to gauge 
theories that have structure functions rather than constants (soft algebras), 
and
to the case where the gauge transformations may or may not be independent, 
reducible or irreducible algebras respectively.
Zinn-Justin introduced the concept of sources of BRST-transformations 
\cite{ZJ}.  These sources are the antifields in the BV formalism.  It was 
shown also that the geometry of the antifields have a natural origin 
\cite{Wit}.

At the quantum level, the field-antifield formalism also works at one-loop 
anomalies \cite{Troost,Pro}.  There, with the addition of extra degrees of 
freedom, which origins an extension of the original configuration space, we
have 
a solution for the regularized quantum master equation (QME) at one-loop,
that has 
been obtained as an independent part of the antifields inside the anomaly.

Let us construct the complete set of fields, including in this set 
the classical fields, the
ghosts for all gauge symmetries and the auxiliary fields.  This
complete set will be denoted by ${\Phi^{A}}$.  Now, let us extend this
space with the same number of fields, but at this time, one will
define the antifields ${\Phi_{A}^{*}}$, which is the canonical
conjugated variables with respect to the antibracket structure.  This
can be written as
\beq
(X,Y) = \frac{\delta_{r} X}{\delta \phi}\,\frac{\delta_{l} Y}{\delta
\phi^{*}} - ( X \longleftrightarrow Y ),
\eeq
where the indices $r$ and $l$ denote right and left derivation
respectively. 

Concerning the antibrackets, one can write the canonical conjugation
relations 
\beq
\label{conj}
(\Phi^{A},\Phi^{*}_{A}) = \delta^{A}_{B}\,\,,\,\,\,
(\Phi^{A},\Phi^{B}) = (\Phi^{*}_{A},\Phi^{*}_{B}) = 0.
\eeq

The antifields $\Phi^{*}_{A}$ have opposite statistics than their
conjugated fields $\Phi^{A}$.  The antibracket is a fermionic operation
so that the statistics of the antibracket $(X,Y)$ is opposite to that
of $XY$.  The antibracket also satisfies some graded Jacobi relations,
\beq
(X,(Y,Z)) +(-)^{\epsilon_{X}\epsilon_{Y} + \epsilon_{X} +
\epsilon_{Y}} (Y,(X,Z)) = ((X,Y),Z).
\eeq
where $\epsilon_{X}$ is the statistics of $X$, i.e. 
$\epsilon(X) = \epsilon_{X}$.

We will define a quantity named ghost number for fields and antifields.
These are integers numbers such that
\beq
gh(\Phi^{*}) = - 1 - gh(\Phi).
\eeq

One can then construct an action of ghost number zero so that it is an
extended action, the so called BV action, also called classical
proper solution, so that
\beq
\label{proaction}
S(\Phi,\Phi^{*}) = S_{0}(\Phi) + \Phi^{*}_{A} R^{A}(\Phi) +
\frac{1}{2}\Phi^{*}_{A} \Phi^{*}_{B} R^{AB}(\Phi) + \ldots +
\frac{1}{n!} \Phi^{*}_{A_{1}} \ldots \Phi^{*}_{A_{n}} R^{A_{n}\ldots
A_{1}} + \ldots
\eeq

This equation contains the whole algebra of the theory, the gauge
invariance of the classical action $(S_{cl} =
S_{BV}(\Phi^{A},\Phi^{*}_{A}=0))$, Jacobi identities,...
Gauge fixing is obtained either by a canonical transformation or by
choosing a fermion $\Psi^{A}$ and writing 
\beq
\Phi^{*} = \frac{\delta_{r} \Psi^{A}}{\delta \Phi^{A}}
\eeq

At the quantum level the quantum action can be defined by
\beq
\label{exp}
W = S + \sum^{\infty}_{p=1} \hbar^{p} M_{p},
\eeq
where the $M_{i}$ are the quantum corrections, the Wess-Zumino terms,
to the quantum action.  The expansion (\ref{exp}) is not the only one, 
but is
the usual one.  An expansion in $\sqrt{\hbar}$ \cite{Sie}  can be made.
This will originate the so called background charges, which is useful
in conformal field theory \cite{Back}.

The quantization of the theory is made by the Green function's
generating functional
\beq
Z(J,\Phi^{*}) = \int {\cal D} \Phi \,exp \, \frac{i}{\hbar} \left[ W (\Phi,
\Phi^{*}) + J_{A} \Phi^{*A} \right] .
\eeq
Concerning the regularization framework, the definition of a path integral is missing, which can be seen as a way to define the
measure. Anomalies represent the non conservation of classical
symmetries at quantum level.  
%Physically when the theory has an
%anomaly, the antifields become propagating fields.
%These are called induced theories.

For a theory to be free of anomalies, the quantum action $W$ has to
be a solution of the QME,
\beq
\label{qme}
(W,W) = 2\,i\,\hbar\, \Delta\, W
\eeq
where
\beq
\Delta \equiv (-1)^{A+1} \frac{\partial_{r}}{\partial \Phi^{A}}
\frac{\partial_{r}}{\partial\Phi^{*}_{A}}.
\eeq

In Eq. (\ref{qme})  one can see that:
\beq
{\cal A} \equiv \left[ \Delta W + \frac{i}{2\hbar} (W,W) \right] (\Phi, \Phi^{*}).
\eeq
And computing a $\hbar$ expansion,
\beq
{\cal A} = \sum^{\infty}_{p=0} \hbar^{p-1} M_{p}
\eeq
one have the form of the $p$-loop BRST anomalies,
\ben
{\cal A}_{0} & = & \frac{1}{2}\,(S,S) \equiv 0 \\
{\cal A}_{1} & = & \Delta S + i\,(M_{1},S) \\
{\cal A}_{2} & = & \Delta M_{p-1} +
\frac{i}{2}\sum^{p-1}_{q=1} (M_{q},M_{p-q}) + i(M_{p},S)\,\,, p \geq 2
\een

The first equation is the known CME.  The second one is an equation
for $M_{1}$.  If the second equation does not have a solution for
$M_{1}$ then ${\cal A}$ is called anomaly.  The anomaly is not
uniquely determined since $M_{1}$ is arbitrary. and it satisfies
the Wess-Zumino consistency condition \cite{WZ},
$({\cal A},S) = 0$.

%If we split the equation (\ref{qme}) in powers of $\hbar$, the two first
%equations, the zero-loop and the one-loop are:
%\ben
%(S,S) & = & 0 \\
%i {\cal A} & \equiv & i \Delta S - (M_{1},S) = 0.
%\een

\subsection{The Non-local Regularization}

When the Wess-Zumino terms, which cancel the anomaly,
can not be found, the theory can be said to have a 
genuine anomaly.  A few years back, a method was developed to handle with global 
anomalies \cite{Nelson}, i.e., when a quantity that is conserved classically is 
not conserved at quantum level.

However, the solution of the QME is not easily obtained because there is a 
divergence when the $\Delta$ operator, a two order differential operator 
defined below, is applied on local functionals, 
a $\delta(0)$-like divergence.  Therefore, a 
regularization method has to be used to cut the divergence
 in the QME.  One of these methods is the well known Pauli-Villars (PV) regularization 
\cite{Pauli,Diaz,Hat}, where new fields, the PV fields, and an arbitrary mass 
matrix are introduced.  However, this method is very useful only at one-loop 
level.  At higher orders, the PV method is still mysterious.  The 
BPHZ renormalization \cite{BPHZ} of the BV formalism was formulated 
\cite{Jonghe,EU}.  A dimensional regularization method in the quantum aspect 
of the field-antifield quantization has been studied in ref. \cite{Tonin}.

The non-local regularization (NLR) \cite{NL,Kle,Woo} gives a consistent way to 
compute anomalies at higher order levels of $\hbar$.  The main ideas were 
based on Schwinger's proper time method \cite{Sch}.  The 
preliminary results \cite{Hand,Clay} were very well received. The  
NLR separates the 
original divergent loop integrals in a sum over loop contribution in such a 
way that the loops, now composed of a set of auxiliary fields, contain the 
original singularities.  To regularize the original theory one has to 
eliminate these auxiliary fields by putting them on shell.  In this way the 
theory is free of the quantum fluctuations.  An extension of the NLR
method to the BV framework has been formulated in \cite{Paris}.

As we explained at the introduction, the non-local regularization can
be applied only to theories which have a perturbative expansion,
i.e. for actions that can be decomposed into a free and an
interacting part.  For much more details, including the diagrammatic
part, the interested reader can see the references \cite{NL,Kle,Woo,Paris}.

Let us define an action $S(\Phi)$ where $\Phi$ is the set $\Phi^{A}$
of the fields, $A=1,\ldots,N$, and with statistics
$\epsilon(\Phi^{A}) \equiv \epsilon_{A}$.
\beq
\label{origaction}
S(\Phi) = F(\Phi) + I(\Phi),
\eeq
$F(\Phi)$ is the kinetic part and $I(\Phi)$ is the interacting part.
$F(\Phi)$ can be written as
$F(\Phi) = \frac{1}{2} \Phi^{A} {\cal F}_{AB} \Phi^{B}$
and $I(\Phi)$ is an analytic function in $\Phi^{A}$ around
$\Phi^{A} = 0$. ${\cal F}_{AB}$ is called the kinetic
operator. 

We have now to introduce a cut-off or regulating parameter
$\Lambda^{2}$.  An arbitrary and invertible matrix $T_{AB}$ has to
be introduced too.  With the combination between ${\cal F}_{AB}$ and 
$(T^{-1})^{AB}$ we can define a second order derivative regulator
\beq
{\cal R}^{A}_{B} = (T^{-1})^{AC} {\cal F}_{AB}.
\eeq
which will help in the construction of two important operators.  The first one is the smearing operator
\beq
\epsilon^{A}_{B} = exp \left( \frac{{\cal R}^{A}_{B}}{2\Lambda^{2}} \right) ,
\eeq
and the second one is the shadow kinetic operator
\beq
{\cal O}_{AB}^{-1} = T_{AC}(\tilde{{\cal O}}^{-1})^{C}_{B} = 
\left( \frac{{\cal F}}{\epsilon^{2} - 1} \right)_{AB},
\eeq
where $(\tilde{{\cal O}})^{A}_{B}$ is defined as
\beq
\tilde{\cal O}^{A}_{B} = \left( \frac{\epsilon^{2} - 1}{{\cal R}} \right)^{A}_{B}
= \int_{0}^{1} \frac{dt}{\Lambda^{2}}\, 
exp \left(\,t\, \frac{{\cal R}^{A}_{B}}{\Lambda^{2}} \right).
\eeq

For each field $\Phi^{A}$ an auxiliary field $\Psi^{A}$ can be constructed, i.e., the
shadow field, with the same statistics.  A new auxiliary action
couple both sets of fields
\beq
\label{auxaction}
\tilde{{\cal S}}(\Phi,\Psi) = F(\hat{\Phi}) - A(\Psi) + I(\Phi + \Psi).
\eeq
The second term of this auxiliary action is called kinetic term, 
\beq
A(\Psi) = \frac{1}{2}\Psi^{A} ({\cal O}^{-1})_{AB} \Psi^{B}.
\eeq
The fields $\hat{\Phi}^{A}$, the smeared fields, which make part of
the auxiliary action are defined by
\beq
\hat{\Phi}^{A} \equiv (\epsilon^{-1})^{A}_{B} \Phi^{B}.
\eeq

It can be proved that, to eliminate the quantum fluctuations
associated with the shadow fields at the path integral level one has to
accomplish this by putting the auxiliary fields $\Psi$ on shell.  So, the
classical shadow fields equations of motion are
\beq
\label{claeq}
\frac{\partial_{r} \tilde{S}(\Phi,\Psi)}{\partial \Psi} = 0
\Longrightarrow \Psi^{A} =\left( \frac{\partial_{r} I}{\partial
\Phi^{B}}(\Phi + \Psi) \right) {\cal O}^{BA}.
\eeq
These equations can be solved in a perturbative fashion.  The
classical solutions $\bar{\Psi}_{0}(\Phi)$ can now be substituted in
the auxiliary action (\ref{auxaction}).  This substitution modify the
auxiliary action so that a new action, the non-localized action appear,
\beq
\label{nlocaction}
{\cal S}_{\Lambda}(\Phi) \equiv \tilde{\cal S}(\Phi,\bar{\Psi}_{0}(\Phi))\,\,,
\eeq
which can be expanded in $\bar{\Psi}_{0}$.
As a result, we see the appearance of the smeared kinetic term
$F(\hat{\Phi})$, the original interaction term $I(\Phi)$ and an
infinite series of new non-local interaction terms.  But all these
interaction terms are $O\left(\frac{1}{\Lambda^{2}}\right)$.
When the limit $\Lambda^{2} \longrightarrow \infty$ is taken, we
will have that ${\cal S}_{\Lambda}(\Phi) \longrightarrow {\cal
S}(\Phi)$, and the original theory is recovered.  Equivalently to this
limit, the same result can be obtained with the limits
\beq
\epsilon \longrightarrow 1,\,\,\,\,\,\,\,\,\,\, {\cal O} \longrightarrow 0,
\,\,\,\,\,\,\,\,\,\, \bar{\Psi}_{0}(\Phi) \longrightarrow 0.
\eeq

With all this framework, when we introduce the smearing operator, any
local quantum field theory can be made ultraviolet finite.  But a
question about symmetry can appear.   Obviously this form of
non-localization destroy any kind of gauge symmetry or its associated
BRST symmetry.  The final consequence is the damage of the
corresponding Ward identities at the tree level.
If the original action (\ref{origaction}) is invariant under the infinitesimal
transformation 
\beq
%\label{brst}
\delta\,\Phi^{A} = R^{A}(\Phi)\,\,,
\eeq
the auxiliary action is invariant under the auxiliary
infinitesimal transformations
\ben
\label{brst}
\tilde{\delta} \Phi^{A} & = & \left( \epsilon^{2} \right)^{A}_{B}\,R^{B}\,
(\Phi + \Psi), \nonumber \\
\tilde{\delta} \Psi^{A} & = & \left( 1-\epsilon^{2} \right)^{A}_{B}\,
R^{B}\,(\Phi + \Psi).
\een

However, the non-locally regulated action (\ref{nlocaction}) is
invariant under the transformation
\beq
\delta_{\Lambda} (\Phi^{A}) = \left( \epsilon^{2} \right)^{A}_{B}\,R^{B} \left( \Phi + 
\bar{\Psi}_{0}(\Phi) \right),
\eeq
where $\bar{\Psi}_{0}(\Phi)$ are the solutions of the
classical equations of motions (\ref{claeq}).

Hence, any of the original continuous symmetries of the theory are
preserved at the tree level, even the BRST transformations, and
consequently, the
original gauge symmetry.  The reader can see \cite{NL,Kle,Woo} for
details.

\subsection{The Extended (BV) Non-local Regularization}

Using the construction, described in the last section, of the NLR and the BV
results, one can build a regulated BRST classical structure of a
general gauge theory from the original one.  Consequently, a
non-locally regularized BV formalism comes out.

The BV configuration space has
to be enlarged introducing the antifields $\{\Psi^{A},\Psi_{A}^{*}\}$.
Note that the shadow fields have antifields too.  Then, an auxiliary
proper solution, which incorporates the auxiliary
action (\ref{auxaction}), corresponding to the gauge-fixed action
$S(\Phi)$, its BRST symmetry (\ref{brst}) and the unknown
associated higher order structure functions.  The auxiliary BRST
transformations (\ref{brst}), are modified by the presence of the
term $\Phi^{*}_{A}\,R^{A}(\Phi)$ in the original proper solution.
Then it can be written that the BRST transformations are 
\beq
\left[ \Phi^{*}_{A}(\epsilon^{2})^{A}_{B} +
\Psi^{*}_{A}(1-\epsilon^{2})^{A}_{B} \right] \,R^{B}\,\left(\Phi +
\Psi \right)
\eeq
which are originated from the substitution
\ben
\label{subst}
\Phi^{*}_{A} & \longrightarrow & \left[ \Phi^{*}_{A}(\epsilon^{2})^{A}_{B} +
\Psi^{*}_{A}(1-\epsilon^{2})^{A}_{B} \right] \equiv \Theta^{*}_{A}
\nonumber \\
R^{A} & \longrightarrow &R^{A}(\Phi + \Psi) \equiv R^{A}(\Theta).
\een

For higher orders, the natural way would be
\beq
R^{A_{n}\ldots A_{1}}(\Phi) \longrightarrow
R^{A_{n}\ldots A_{1}}(\Phi+\Psi) = R^{A_{n}\ldots A_{1}}(\Theta)
\eeq
and an obvious ansatz for the auxiliary proper solution is
\ben
\label{ansataction}
\tilde{S}(\Phi,\Phi^{*};\Psi,\Psi^{*}) \, & = & \,\tilde{S}(\Phi,\Psi) \,+\,
\Theta^{*}_{A}\,R^{A}(\Theta) \, + \,
\Theta^{*}_{A}\Theta^{*}_{B}\,R^{AB}(\Theta) \, +  \nonumber \\
& + & \Theta^{*}_{A_{1}}\ldots \Theta^{*}_{A_{n}}\,R^{A_{n}\ldots A_{1}}(\Phi)
+ \ldots
\een

It is intuitive to see that the same canonical conjugation relations, 
equations (\ref{conj}), should be obtained, i.e.
\beq
\left( \Theta^{A},\Theta^{*}_{B} \right) = \delta^{A}_{B}.
\eeq
Consequently, we have to construct a new set of fields and antifields
$\{\Sigma^{A},\Sigma^{*}_{A}\}$ defined by
\beq
\Sigma^{A} = \left[ \left( 1-\epsilon^{2} \right)^{A}_{B}\Phi^{B} -
\left( \epsilon^{2} \right)^{A}_{B}\Psi^{B} \right],
\eeq
and
\beq
\Sigma^{*}_{A} = \Phi^{*}_{A} - \Psi^{*}_{A}.
\eeq

Now we have that the linear transformation
\beq
\{\Phi^{A},\Phi^{*}_{A};\Psi^{A},\Psi^{*}_{A}\} \longrightarrow 
\{\Theta^{A},\Theta^{*}_{A};\Sigma,\Sigma^{*}_{A}\}
\eeq
is canonical in the antibracket sense.  And the auxiliary
action (\ref{auxaction}) is the original proper
solution (\ref{proaction}) with arguments
$\{\Theta^{A},\Theta^{*}_{A}\}$.

The elimination of the auxiliaries fields of BV method is the next
step.  The shadow fields have to be substituted by the solutions of
their classical equations of motion.  At the same time, their
antifields goes to zero.  In this way we can write that
\beq
S_{\Lambda}(\Phi,\Phi^{*}) =
\tilde{S}(\Phi,\Phi^{*};\bar{\Psi},\Psi^{*} = 0),
\eeq
and the classical equations of motion are
\beq
\frac{\delta_{r}\,\tilde{S}(\Phi,\Phi^{*};\Psi,\Psi^{*})}
{\delta\,\Psi^{A}} = 0
\eeq
with solutions $\bar{\Psi} \equiv \bar{\Psi}(\Phi,\Phi^{*})$, which
explicitly read
\beq
\label{solution}
\bar{\Psi}^{A} = \left[ \frac{\delta_{r}\,I}{\delta \Phi^{B}}\,
\left(\Phi + \Psi \right)
+ \Phi^{*}_{C} \left( \epsilon^{2} \right)^{C}_{D} R^{D}_{B} 
\left( \Phi+\Psi \right) +
O \left( (\Phi^{*})^{2} \right) \right]
\eeq
with
\beq
R^{A}_{B} = \frac{\delta_{r}\,R^{A}\,(\Phi)}{\delta \Phi^{B}}.
\eeq
The lowest order of equation (\ref{solution}) is,
\beq
\bar{\Psi}^{A} = \left( \frac{\delta_{r}\,I}{\delta \Phi^{B}}(\Phi + \Psi)
\right) {\cal O}^{BA}
\eeq
and one can obtain an expression for $\bar{\Psi}(\Phi,\Phi^{*})$ at
any desired order in antifields \cite{Paris}.

To quantize the theory, it is necessary to add extra counterterms
$M_{p}$ to preserve the quantum counterpart of the classical BRST
scheme.  It is the same as to substitute the classical action $S$ by
a quantum action $W$.  
%In the original papers \cite{NL,Kle,Woo} 
%the quantization of the theory
%was already analyzed, but it seems that only one-loop $M_{1}$ 
%corrections acquired
%BRST invariance.  
It can be proved that in the field-antifield
framework, in general, two and higher order loop corrections should
also be considered \cite{Paris}.

The complete interaction term ${\cal I}(\Phi,\Phi^{*})$ of the original
proper solution can be written as
\beq
{\cal I}(\Phi,\Phi^{*}) \equiv I(\Phi) + \Phi^{*}_{A}\,R^{A}(\Phi) +
\Phi^{*}_{A}\,\Phi^{*}_{B}\,R^{AB}(\Phi) + \dots
\eeq
The non-localization of this interaction part furnishes a way to
regularize interactions from counterterms $M_{p}$.  To construct the
auxiliary free and interaction parts we have that
\beq
\tilde{F}\,(\Phi + \Psi) = F(\hat{\Phi}) - A(\Psi),\,\,\,\,\,\,{\cal
I}\,(\Phi,\Phi^{*};\Psi,\Psi^{*}) = {\cal I}\,(\Theta,\Theta^{*})
\eeq
with $\{\Theta,\Theta^{*}\}$ already known.

Now one have to put the auxiliary fields on shell and its
antifields to zero, so that
\ben
F_{\Lambda}\,(\Phi,\Phi^{*}) & = & \tilde{F}\,(\Phi,\bar{\Psi}_{0}),
\nonumber \\
{\cal I}_{\Lambda}(\Phi,\Phi^{*}) & = & \tilde{{\cal
I}}\,(\Phi+\bar{\Psi}_{0},\Phi^{*} \epsilon^{2}),
\een
then $S_{\Lambda}=F_{\Lambda} + {\cal I}_{\Lambda}$.

The quantum action $W$ can be expressed by
\beq
W = F + {\cal I} + \sum_{p=1}^{\infty}\,\hbar\,M_{p} \equiv F + {\cal
Y}
\eeq
where ${\cal Y}$ now is the generalized quantum interaction.
%An analogous procedure of the previous section can be applied to the quantum action $W$.  We will omit all the formal steps here.
A decomposition in its divergent part and its finite part when
$\Lambda^{2} \longrightarrow \infty$ can be accomplished in the
regulated QME.

It can be shown that the expression of the anomaly is the value of
the finite part in the limit $\Lambda^{2} \longrightarrow \infty$ of 
\beq
\label{anomalia}
{\cal A} = \left[ (\,\Delta\,W\,)_{R} + \frac{i}{2\,\hbar}\,(W,W)
\right]\,(\Phi,\Phi^{*})
\eeq
and the regularized value of $\Delta W$ defined as
\beq
\label{operator}
(\Delta W)_{R} \equiv \lim_{\Lambda^{2} \rightarrow \infty} \left[
\Omega_{0} \right]
\eeq
where
\beq
\Omega_{0} = \left[
S_{B}^{A}\,\left( \delta_{\Lambda} \right)^{B}_{C}\, \left( \epsilon^{2} \right)_{A}^{C} \right].
\eeq

\noindent and $\left( \delta_{\Lambda} \right)^{A}_{B}$ is defined by
\ben
(\delta_{\Lambda})_{B}^{A} & = & \left( \delta^{A}_{B} - {\cal
O}^{AC}\,{\cal I}_{CB} \right)^{-1} \nonumber \\
& = & \delta^{A}_{B} + \sum_{n=1}\, \left( {\cal O}^{AC}\,{\cal
I}_{CB} \right)^{n},
\een
with
\ben
S^{A}_{B} & = &
\frac{\delta_{r}\,\delta_{l}\,S}{\delta\,\Phi^{B}\,\delta\,\Phi^{*}_{A}}, \nonumber \\
{\cal I}_{AB} & = &
\frac{\delta_{r}\,\delta_{l}\,{\cal I}}{\delta\,\Phi^{A}\,\delta\,\Phi^{B}} 
\een

Applying the limit $\Lambda^{2} \longrightarrow \infty$ in 
(\ref{operator}), it can be shown that
\beq
\left( \Delta S\right)_{R} \equiv \lim_{\Lambda^{2} \rightarrow \infty}
\left[ \Omega_{0} \right]_{0}
\eeq

And finally that
\ben
{\cal A}_{0} & \equiv & \left( \Delta\,S \right)_{R} \nonumber \\
& = & \lim_{\Lambda^{2} \rightarrow \infty} \left[ \Omega_{0}
\right]_{0} 
\een

All the higher orders loop terms of the anomaly can be obtained from
equation (\ref{anomalia}), but this will not be analyzed in this paper.

\section{The NC anomaly of the CSM}

We know that the CSM is a $U(1)$ gauge field coupled to chiral fermions and in spite of being anomalous, constitute a consistent unitary theory.  Moreover, the fermionic determinant and the anomaly have some arbitrariness relative to the regularization of fermionic radiative contributions (see \cite{JR} for a review).

The classical action for the chiral Schwinger model is
\beq
\label{csmaction}
S = \int\,d^{2}x \left[  - \frac{1}{4} \, F_{\mu\nu} F^{\mu\nu} +
\bar{\psi} i \not\partial \psi + {e \over 2}\bar{\psi}\gamma_{\mu} (1 -
\gamma_{5}) A^{\mu} \psi \right],
\eeq
which obviously has a perturbative expansion.

This action is invariant for the following  gauge transformations:
\ben
A^{\mu}(x) & \longrightarrow & A^{\mu}(x) + \partial_{\mu} \theta (x) \\
\psi(x) & \longrightarrow & exp\, \left[
i\,e\,(1-\gamma_{5})\,\theta(x)\, \right]\,\psi(x)
\een

The kinetic part of the action (\ref{csmaction}) is given by
\ben
F & = & \int d^{2}x\,\bar{\psi}\,i\not\partial \psi  \nonumber \\
& = & \int d^{2}x\, \left[ \frac{1}{2} \bar{\psi} \,i\not\partial \psi + 
\frac{1}{2} \bar{\psi} \,i \not\partial \psi \right]
\een

Integrating by parts the second term we have that
\beq
F = \int d^{2}x\, \left[ \frac{1}{2}\bar{\psi}\,i\not\partial \psi - 
\frac{1}{2}(i\not\partial^{t}\bar{\psi})\,\psi \right]
\eeq

The kinetic term has the form
\beq
F = \frac{1}{2} \Psi^{A}{\cal F}_{AB}\Psi^{B}\,\,,
\eeq

\noindent where,
$
\Psi = \left( \begin{array}{c}
\bar{\psi} \\ \psi
\end{array} \right)
$
and
\beq
F = \frac{1}{2}(\bar{\psi} \,\,\psi)\, 
\left( \begin{array}{cc}
0 & i\not\partial \\
i\not\partial^{t} & 0 
\end{array} \right) 
\left( \begin{array}{c}
\bar{\psi} \\ \psi 
\end{array} \right)
\eeq
and we have that the kinetic operator $( {\cal F}_{AB} )$ is
\beq
{\cal F}_{AB} = \left( \begin{array}{cc}
0 & i\not\partial \\
i\not\partial^{t} & 0 
\end{array} \right) 
\eeq

The regulator, a second order differential operator, is
$
{\cal R}^{\alpha}_{\beta} = (T^{-1})^{\alpha \gamma}{\cal F}_{\gamma \beta},
$
where $T$ is an arbitrary matrix,
and one can make the following choice,
\beq
{\cal R}^{\alpha}_{\beta} = -\,\partial^{2}
\eeq

Using the definition of the smearing operator,
\beq
\epsilon^{A}_{B} = exp \left( \frac{- \partial^{2}}{2 \Lambda^{2}} \right),
\eeq
and the smeared fields are defined by
$
\hat{\Phi}^{A} = (\epsilon^{-1})^{A}_{B}\,\Phi^{B}\,\,.
$

In the NLR scheme the shadow kinetic operator is
\beq
{\cal O}_{\alpha\beta}^{-1} = 
\left( \frac{{\cal F}}{\epsilon^{2} - 1} \right)_{\alpha\beta}
\eeq
then
\beq
{\cal O} = \left( \begin{array}{cc}
0 & -i{\cal O}'\not\partial \\
-i{\cal O}'\not\partial^{t} & 0 
\end{array} \right) 
\eeq
where
\ben
{\cal O}' & = & \frac{\epsilon^{2} - 1}{\not\partial\not\partial^{t}}\, 
 =\,  \int_{0}^{1} \frac{dt}{\Lambda^{2}} exp
\left( t\, \frac{\not\partial^{t} \not\partial}{\Lambda^{2}} \right)\,\,.
\een

The interacting part of the action (\ref{csmaction}) is
\ben
I\,\left[ A_{\mu},\psi,\bar{\psi} \right] & = &
e\,\bar{\psi}\,\gamma_{\mu}(1-\gamma_{5})A^{\mu}\,\psi \\
I \left[ A_{\mu},\psi+\Phi,\bar{\psi}+\bar{\Phi} \right] & = &
e\,( \bar{\psi} + \bar{\Phi} )\, \gamma_{\mu}(1-\gamma_{5})A^{\mu}\,
( \psi + \Phi )
\een
where $\Phi$ are the shadow fields.

The BRST transformations are given by
\ben
\delta A_{\mu} & = & \partial_{\mu}c, \nonumber \\
\delta \psi & = & i ( 1\,-\,\gamma_5 )\psi c, \nonumber \\
\delta\bar{\psi} & = & -\,i\bar{\psi} ( 1\,+\,\gamma_5 )c, \nonumber \\
\delta c & = & 0\,\,.
\een

Substituting (\ref{subst}) where the antifields are 
functions of the auxiliary fields,
\ben
\psi^{*} \longrightarrow \left[ \psi^{*} \epsilon^{2} + \Phi^{*} (1-
\epsilon^{2}) \right] \nonumber \\
\bar{\psi}^{*} \longrightarrow \left[ \bar{\psi}^{*} \epsilon^{2} + 
\bar{\Phi}^{*} (1 - \epsilon^{2}) \right].
\een

The generator of BRST transformations are
\ben
R(\psi) & \longrightarrow & R(\psi + \Phi) = i ( 1\,-\,\gamma_5 )( \psi + \Phi) c \nonumber \\
R(\bar{\psi}) & \longrightarrow & i (\bar{\psi} + \bar{\Phi})( 1\,+\,\gamma_5 )c \nonumber \\
R(c) & = & 0
\een

The non-local auxiliary proper action will be given in general by
$
S_{\Lambda}(\Phi,\Phi^{*}) =
\tilde{S}_{\Lambda}(\Phi,\Phi^{*};\psi_{s},\psi^{*} = 0)
$
where $\psi_{s}$ are the solutions of the classical equations of motion.

After an algebraic manipulation, one can write the non-localized
action as
\ben
\tilde{S}_{\Lambda}(\psi,\psi^{*}) & = & \hat{F}_{\mu\nu} \hat{F}^{\mu\nu} +
\hat{\bar{\psi}} i\not\partial \hat{\psi} + A^{*}_{\mu}\partial^{\mu}c +
\nonumber \\
& + & \frac{e\,(i\not\partial) \left[ \bar{\psi}\gamma^{\mu} (1 -
\gamma_{5}) A_{\mu} \psi \right]}{i\not\partial+e\,\gamma^{\mu} 
(1 -\gamma_{5}) A_{\mu}(\epsilon^{2}-1)}  + \nonumber \\
& + & \frac{i\,\psi^{*}\epsilon^{2}c(-i\not\partial) \left[ (1 -
\gamma_{5}) \psi \right] }{\left[ i\not\partial+e\,\gamma^{\mu} 
(1 -\gamma_{5}) A_{\mu}(\epsilon^{2}-1) \right]}\, + \nonumber \\
& + & \frac{i\,\bar{\psi}^{*}\epsilon^{2}c(i\not\partial)\left[ \bar{\psi}
(1 -\gamma_{5}) \right]}{ \left[ i\not\partial+e\,\gamma^{\mu} 
(1 -\gamma_{5}) A_{\mu}(\epsilon^{2}-1) \right]}.
\een
It can be seen easily that when one takes the limit 
$\epsilon^{2} \longrightarrow 1$, the original proper solution of the CSM,
shown below, is obtained.

Now we have to construct some very important matrices,
\beq
S_{A}^{B} = \frac{\delta_{r}\delta_{l}\,S_{BV}}{\delta
\Phi^{B}\,\delta \Phi^{*}_{A}}
\eeq
with the proper solution, the BV action, given by
\ben
S_{BV} &=& \int\,d^{2}x \, \{ \; -  \frac{1}{4} \, F_{\mu\nu} F^{\mu\nu} +
\bar{\psi} i \not\partial \psi + {e \over 2}\bar{\psi}\gamma_{\mu} (1 - \gamma_{5}) 
A^{\mu} \psi  +  A^{*}_{\mu}\partial^{\mu}c  \nonumber \\
&+& i\,\psi^{*}(1 -\gamma_{5})\psi c - i\,\bar{\psi}^{*}\bar{\psi}(1 +\gamma_{5})c \;\}
\een
then
\beq
S_{B}^{A} = \left( \begin{array}{cc}
-ic(1 -\gamma_{5}) & 0 \\
0 & ic(1 +\gamma_{5})
\end{array} \right) .
\eeq

The operator ${\cal I}_{AB}$ in this case is,
\beq
{\cal I}_{AB} = \frac{\delta_{l}\delta_{r} \left[ I(\Phi) +
\Phi^{*}_{c}R^{c}(\Phi) \right]}{\delta \Phi^{A}\delta \Phi^{B}}
\eeq
and the result is,
\beq
{\cal I}_{AB} = \left( \begin{array}{cc}
0 & -\,{e \over 2}\gamma_{\mu} (1 -\gamma_{5}) A^{\mu} \\
{e \over 2}\gamma_{\mu} (1 -\gamma_{5}) A^{\mu} & 0
\end{array} \right) 
\eeq

The one-loop anomaly is given by:
\ben
{\cal A} & \equiv & (\Delta S)_{R} \\
(\Delta S)_{R} & = & \lim_{\Lambda^{2} \rightarrow \infty}
[\Omega_{0}]_{0} \\
\Omega_{0} & = & \left[ \epsilon^{2}S_{A}^{A} \right] +
\left[ \epsilon^{2}S_{B}^{A}{\cal O}^{BC}{\cal I}_{CA} \right] + 
O \left( \frac{(\Phi^{*})^{2}}{\Lambda^{2}} \right)
\een

For the first term
\ben
\epsilon^{2}S_{A}^{A} & = & \epsilon^{2}\,tr\,S_{B}^{A} \nonumber \\
& = & 0
\een
and we have that
\beq
(\Delta S)_{R} = \lim_{\Lambda^{2} \rightarrow \infty} 
tr \left[ \epsilon^{2}S_{B}^{A}{\cal O}^{BC}{\cal I}_{CA} \right]
\eeq

Using the $\gamma$ matrix representation
\beq
\gamma^{0} = \left( \begin{array}{cc}
0 & -1 \\
-1 & 0
\end{array} \right)         
\eeq
\beq
\gamma^{1} = \left( \begin{array}{cc}
0 & -i \\
i & 0
\end{array} \right)         
\eeq
and
\beq
\gamma^{5} = \,-\,i\,\gamma_{1}\,\gamma_{0}
\eeq
in this representation we have that $\gamma_{5}^t=\gamma_{5}$.

Finally, we have that,
\beq
(\Delta S)_{R} = \lim_{\Lambda^{2} \rightarrow \infty} 
tr \left[ \epsilon^{2}(-ec)\frac{\epsilon^{2}-1}{\partial^{2}}
(\partial_{\mu}A^{\mu}\,-\,\epsilon^{\mu\nu} \partial_{\mu}A_{\nu} ) \right]\,\,.
\eeq

But we know that
\ben
& & \lim_{\Lambda^{2} \rightarrow \infty} 
tr \left[ \epsilon^{2}\, F \partial^{n} \, \frac{\epsilon^{2}-1}{\partial^{2}}\,
\partial \, G \, \partial^{m} \right] = \\
& = & \frac{-i}{2 \pi} \left[ \, \sum_{k=0}^{m}
\left( \begin{array}{c}
m \\ k
\end{array} \right)\frac{(-1)^{k}}{n+m+1-k} 
\left( 1-\frac{1}{2^{n+m+1-k}} \right) \right]
\int \, d^{2}x\,F\,\partial^{n+m+1}\,G \,\,.\nonumber
\een    

In our case
\ben
n & = & m = 0 \nonumber \\
F & = & 2ec \nonumber \\
\partial G & = & \partial_{\mu}A^{\mu}\,-\,\epsilon^{\mu\nu} \partial_{\mu}A_{\nu}   
\een
and 
%the final result is
%\beq
%{\cal A} = (\Delta S)_{R} = \frac{ie}{2 \pi} 
%\int\,d^{2}x\,c\, \left( \partial_{\mu}A^{\mu}\,-\,\epsilon^{\mu\nu} \partial_{\mu}A_{\nu} \right)
%\eeq
%which is the one-loop anomaly of the CSM.
%Now, 
with Eq. (\ref{a7777}) we can construct the NC version of the CSM, which can be written as,

\beq
\label{NCanomaly}
\hat{{\cal A}}\, =\,\frac{ie}{2\pi}\left\{\int d^2 x\, c \left( \partial_{\mu}A^{\mu}\,-\,\epsilon^{\mu\nu} \partial_{\mu}A_{\nu} \right)\,+\,\theta^{\alpha\beta}\partial_{\alpha}\left[A_{\beta}\int d^2 x c \left( \partial_{\mu}A^{\mu}\,-\,\epsilon^{\mu\nu} \partial_{\mu}A_{\nu} \right)\right]\right\}\,+\,O(\theta^2)
\eeq

\noindent where in two dimensions we have that $\theta^{01}=\theta{10}=\theta$.
The first term is the one loop ordinary anomaly of the CSM, i.e., ${\cal A} = (\Delta S)_{R}$, and the second term, of course, is the NC correction term.  To investigate under what conditions the NC anomaly cancels is beyond the scope of this work.  We can see easily form (\ref{NCanomaly}) that, like in NC QED, the anomaly shares terms with the commutative primary theory.  It would be interesting to calculate the $\theta$-second order terms.

\section{Conclusions}

One of the greatest motivations to study noncommutativity is the introduction of a minimal length scale, which is one of the ingredients of quantum gravity.  The noncommutation between the coordinates and momenta in quantum theory insinuates naturally that the same behavior can be performed by the coordinates.  This would cause the introduction of a new scale in the theory through the NC parameter, which could be used, for example, to tame the divergences in QFT.  Nevertheless, the renormalization procedures showed great success and the NC scenarios were put to sleep for more than fifty years until string theory bring them back recently.

Since the NC space can be understood as a deformation of the ordinary spacetime, one consequence of NC effects in QFT is to introduce a combination of IR and UV divergences through the appearing of phase factors in the vertice.

Concerning anomalies in NC scenarios, it was noticed that, resulting only from noncommutativity, two different currents can be defined even for a $U(1)$ theory \cite{as}.  Another example is the description of the $\theta$-structure of the commutator anomalies in NC electrodynamics \cite{bk1}.  So, the treatment of NC anomalies can be considered a quite nontrivial issue and deserves more investigations.

There are two ways to accomplish NC anomalies.  One of them is to introduce noncommutativity in the primary theory and after that to compute the anomaly \cite{huffel,bg}.  The second one is to use the SW map directly in the ordinary result of the anomaly, which was accomplished here for the first time.

In this work we decided in favour of the second one and we used the BV quantization to calculate the CSM anomaly.  The regularization scheme used here was the non-local regularization formalism.  The field-antifield framework exhibits a divergence on the application of the $\Delta$ operator and hence it needs a regularization. 
This is a recent and a quite powerful method to regularize theories with a perturbative expansion which
have higher loop order divergences.  This arguably makes this method the ideal one to analyze NC theories with higher orders in $\theta$.

%The connection between both generates an extended non-locally regularized  BV quantization method.  The quantization of anomalous gauge theories can be computed exactly.  
%The one-loop anomaly of the chiral Schwinger model has been calculated.

The anomaly in (\ref{NCanomaly}) shows that it would be interesting to calculate superior orders in the $\theta$-parameter in order to see if higher terms shares something with the ordinary anomaly.   Hence, as a direct perspective we can develop the SW map, concerning the anomaly, to calculate a general form for superior orders terms for the anomaly.  This is a work in progress.

\vspace{1cm}

\noindent {\bf Acknowledgment:} The authors would like to thank Conselho Nacional de Desenvolvimento Cient\' ifico e Tecnol\'ogico (CNPq) and CAPES,  Brazilian Research Agencies, for financial support.

%\newpage

%\vskip 1cm

%\vspace{1cm}


\begin{thebibliography}{30}

\bibitem{iz}   C. Itzykson and J.-B. Zuber, ``Quantum Field Theory," Dover Books on Physics, New York, 2006.

\bibitem{bertlman} R. A. Bertlmann, ``Anomalies in quantum field theory," Oxford University, New York, 2000; K. fujikawa and H. Suzuki, ``Path integrals and quantum anomalies," Oxford University, New York, 2004.

\bibitem{NC} M. R. Douglas and N. A. Nekrasov, Rev. Mod. Phys. 73 (2001) 977; RJ Szabo, Phys. Rept. 378 (2003) 207; R. Banerjee, B. Chakraborty, S. Ghosh, P. Mukherjee and S. Samanta, Found. Phys. 39 (2009) 1297.

\bibitem{sw}   N. Seiberg and E. Witten, JHEP 09 (1999) 032.

\bibitem{BV}   I. A. Batalin and G. A. Vilkovisky, Phys. Lett. B 102 (1981) 27, 
Phys. Rev. D 28 (1983) 2567.

\bibitem{Jon}   F. DeJonghe,``The Batalin-Vilkovisky Lagrangian Quantization 
Scheme with Applications to the Study of Anomalies in Gauge Theories",Ph.D. 
thesis K. U. Leuven, arXiv: hep-th/9403143.

\bibitem{Gomis}   J. Gomis, J. Paris and S. Samuel, Phys. Rep. 259 (1995) 1.

\bibitem{Hen}   M. Henneaux, Nucl. Phys.B (Proc. Suppl.) 18 A (1990) 47.

\bibitem{JR}   R. Jackiw and R. Rajaraman, Phys. Rev. Lett. 54 (1985) 1219.

\bibitem{jmssw}   B. Jurco, L. Moller, S. Schraml, P. Schupp and J. Wess, Eur. Phys. J. C 21 (2001) 383.

\bibitem{uy}   K. \"Ulker and B. Yaspiskan, Phys. Rev. D 77 (2008) 065006.

\bibitem{bk1}    R. Banerjee and K. Kumar, Phys. Rev. 72 (2005) 085012.

\bibitem{bk2}    R. Banerjee and K. Kumar, Phys. Rev. D 71 (2005) 045013.

\bibitem{abj}   S. L. Adler, Phys. Rev. 177 (1969) 2426; J. S. Bell and R. Jackiw, Nuovo Cimento A 60 (1969) 47.


\bibitem{ZJ}   J. Zinn-Justin, in Trends in Elementary Paritcle Theory,
Lecture notes in Physics 37, Int. Summer Inst. on Theor. Phys., Bonn 1974, 
eds. H. Rollnik and K. Dietz (Springer, Berlin, 1975); Nucl. Phys. 
B 246 (1984) 246.


\bibitem{Wit}   E. Witten, Mod. Phys. Lett. A 5 (1990) 487; A. Schwarz,
Commun. Math. Phys. 155 (1983) 249; O. M. Khudaverdian and A. P. Nercessian, 
hep-th 9303136; S. Aoyama and S. Vandoren, arXiv: hep-th/9305087.



\bibitem{Troost}   W. Troost, P. van Nieuwenhuizen and A. van Proyen,
Nucl. Phys. B 333 (1990) 727.

\bibitem{Pro}   A. van Proyen, in Proc. Conf. and Symmetries, 1991, Stony
Brook,  May 20-25, 1991, eds. N. Berkovits et al., Word Scientific,
Singapure, 1992, p. 388.

\bibitem{Sie}   F. DeJonghe, R. Siebelink and W. Troost, Phys. Lett. B
396 (1993) 295. 

\bibitem{Back}   P. Ginsparg, ``Applied Conformal Field Theory,"
Lectures at Les Houches Summer School, 1988.

\bibitem{WZ}   J. Wess and B. Zumino, Phys. Lett. B 37 (1971) 95; W. A.
Bardeen and B. Zumino, Nucl. Phys. B 244 (1984) 421.

\bibitem{Nelson}   R. Amorin and N. R. F. Braga, Phys. Rev. D 57 (1998) 1225.

\bibitem{Pauli}   W. Pauli and F. Villars, Rev. Mod. Phys. 21 (1949) 434.

\bibitem{Diaz}   A. Diaz, W. Troost, P. van Nieuwenhuizen and A. van Proyen, 
Int. J. Mod. Phys. A 4 (198) 3959.

\bibitem{Hat}   M. Hatsuda, W. Troost, P. van Neuwenhuizen and A. van Proyen, 
Nucl. Phys. B 335 (1990) 166.

\bibitem{BPHZ}   For a pedagogical account see: W. Zimmerman, in Lectures on 
Elementary Particles and Quantum Field Theory, eds. S. Deser, M. Grisary and 
H. Pendleton, MIT Press, Brandeis Lectures, 1970;
J. Lowenstein, ``Seminars on Renormalization Theory," Technical 
Report no. 73-068, 1972, University of Pittsburgh;
M. O. C. Gomes, ``Some Applications of Normal Product 
Quantization in Renormalization Perturbation Theory," Ph.D. Thesis, 
University of Pittsburg, 1972.

\bibitem{Jonghe}   F. DeJonghe, J. Paris and W. Troost, Nucl. Phys. B 476 (1996) 559.

\bibitem{EU}   E. M. C. Abreu and N. R. F. Braga, 
Int. J. Mod. Phys. A (1998) 4249.

\bibitem{Tonin}   M. Tonin, Nucl. Phys. B (Proc. Suppl.) 29 (1992) 137.

\bibitem{NL}   D. Evens,J. W. Mofat, G. Kleppe and R. P. Woodard, Phys.
Rev. D 43 (1991) 499.

\bibitem{Kle}   G. Kleppe and R. P. Woodard, Ann. Phys. (NY) 221 (1993) 106.

\bibitem{Woo}   G. Kleppe and R. P. Woodard, Nucl. Phys. B 388 (1992) 81.

\bibitem{Sch}   J. Schwinger, Phys. Rev. 82 (1951) 664.

\bibitem{Hand}   B. J. Hand, Phys. Lett. B 275 (1992) 419.

\bibitem{Clay}   M. A. Clayton, L. Demopoulos and J. W. Moffat, 
Int. J. Mod. Phys. A 9 (1994) 4549.

\bibitem{Paris}   J. Paris, Nucl. Phys. B 450 (1995) 357.




\bibitem{as}   F. Ardalan and N. Sadooghi, Int. J. Mod. Phys. A 16 (2001) 3151.


\bibitem{huffel}  H. Huffel, Acta Phys. Slov. 52 (2002) 247.

\bibitem{bg}   K. Bering and H. Grosse, Eur. Phys. J. C 68 (2010) 313.


\end{thebibliography}
\end{document}